\documentclass[smallextended]{svjour3}     
\smartqed  
\usepackage{graphicx}
 \usepackage{mathptmx}      
\usepackage{latexsym}

 \journalname{Climatic Change}
\begin{document}


\title{Are there basic physical constraints on future anthropogenic emissions of carbon dioxide?}


\author
{Timothy J. Garrett }

\institute{Department of Atmospheric Sciences\\
University of Utah\\
Salt Lake City, USA\\
\email{tim.garrett@utah.edu.}
}


\date{}


\maketitle

\begin{abstract}
Global Circulation Models (GCMs) provide projections for future climate warming using a wide variety of highly sophisticated anthropogenic CO$_2$ emissions scenarios as input, each based on the evolution of four emissions {}``drivers'': population $p$, standard of living $g$, energy productivity (or efficiency) $f$ and energy carbonization $c$ \cite{IPCC_WG32007}. The range of scenarios considered is extremely broad, however, and this is a primary source of forecast uncertainty \cite{Stott2002}.  Here, it is shown both theoretically and observationally how the evolution of the human system can be considered from a surprisingly simple thermodynamic perspective in which it is unnecessary to explicitly model two of the emissions drivers: population and standard of living. Specifically, the human system grows through a self-perpetuating feedback loop in which the consumption rate of primary energy resources stays tied to the historical accumulation of global economic production -- or $p\times g$ -- through a time-independent factor of 9.7$\pm$0.3 milliwatts per inflation-adjusted 1990 US dollar. This important constraint, and the fact that $f$ and $c$ have historically varied rather slowly, points towards substantially narrowed visions of future emissions scenarios for implementation in GCMs.   
\end{abstract}

\section{Introduction}

GCM projections of 21st century climate change use prognostic trajectories for carbon dioxide (CO$_2$) emission fluxes developed by the International Panel on Climate Change (IPCC) Special Report on Emissions Scenarios (SRES) \cite{IPCC_WG32007}. These provide a range of timelines, each designed to
show how a given set of  decisions might correspond to a particular
atmospheric CO$_{2}$ trajectory. SRES models  are highly sophisticated,
and contain numerous interactive components, each designed to reflect
a realistic range of societal dynamic behavior.  

For tractability, IPCC SRES models express the primary drivers
of growth in CO$_{2}$ emissions $E$ in terms of human population $p$, primary energy consumption $a$ and real (or inflation-adjusted) economic production $P$ through  $E=p\times g\times i\times c$
 where $g=P/p$ represents the real economic production per person and
$i=1/f = a/P$ represents the {}``energy intensity'' of real economic production, or alternatively, the inverse of its {}``energy productivity'' {$f$}, and $c=E/a$ the carbonization of the energy supply \cite{Nakicenovic2004}. Expressed in a prognostic form, emissions grow at a rate given by \cite{Nakicenovic2004,Raupach2007} \begin{equation}
\frac{d\ln E}{dt}=\frac{d\ln p}{dt}+\frac{d\ln g}{dt}-\frac{d\ln f}{dt}+\frac{d\ln c}{dt}\label{eq:Kaya-prog}
\end{equation}

Differences among SRES emissions trajectories depend on how society is assumed to manage such issues as population control, energy efficiency, and a switch to non-CO$_2$ emitting energy resources. Currently, the range of possible futures considered is extremely broad.  In fact, uncertainty in the degree of surface warming over the next century is determined as much by the range of SRES scenarios as by climate physics itself \cite{Stott2002}.  

In this paper I propose that by using a straight-forward thermodynamic approach it may be possible to substantially constrain plausible timelines for future anthropogenic CO$_2$ emission rates. 

\section{A thermodynamic growth model}

\subsection{A heat engine}
The starting point is recognition that general thermodynamic laws require that all systems, even those that are living, evolve
through a spontaneous conversion of environmental potential energy
into some less available form, often termed {}``heat''  \cite{Schrodinger1944,deGrootMazur1984,Vermeij1995,Kleidon2004}.

\begin{figure*}
\includegraphics[width=0.7\textwidth]{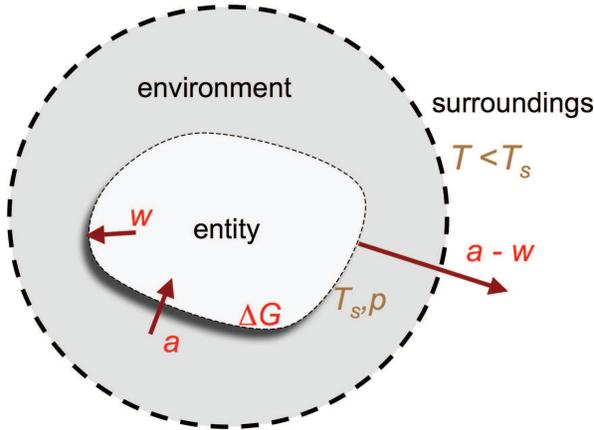}
\caption{\label{fig:thermodynamic}Illustration of an evolving system bounding some entity and its environment, as separated by a permeable interface at constant temperature and pressure. The interface maintains an energy potential $\Delta G\left(T_{s},p\right)$ so that the system as a whole is able to convert available energy at rate $a$ into work $w$ with efficiency $\epsilon = w/a$, and {}``heat''  at rate $a-w$. Heat is voided to the system's colder surroundings; work grows the interface at rate $w=d(\Delta G)/d{t}$.  Because the interface potential is related to energy consumption through $a = \alpha \Delta G$, where $\alpha$ is an engine specific constant coefficient, what is defined is a positive feedback loop in which, through work, $a$ and $\Delta G$ evolve logarithmically at rate $d\ln a/dt = d\ln\left(\Delta G\right)/dt = \eta = \epsilon \alpha$. Here $\eta$ can be considered a feedback efficiency or rate of return.}  
\end{figure*}

Specifically, consider a system drawn in Fig. \ref{fig:thermodynamic}  consisting of some entity and its environment, separated by some arbitrarily defined permeable interface at a fixed temperature $T_{s}$ and pressure $p$ (i.e., at constant energy density or an isentrope). As a whole, the system is in contact (through radiation, convection or conduction) with colder, lower energy density surroundings at temperature $T<T_s$. The interface between the entity and the environment represents a {}``step'', with its height represented by a Gibbs energy potential $\Delta G$. Available potential energy in the environment is converted at rate $a=\alpha\Delta G$ into some less available form through the transfer of matter across the interface. The system-specific constant coefficient $\alpha$ is an intensive quantity that defines the particular physics of {}``availability'' for the system \footnote{To take an electrostatic analogue, $\Delta G$ is a voltage difference, $a$ a current, and $1/\alpha$ is the resistance, in which case the relevant physics defining {}``availability'' of energy is the material's conductivity.}. 

Effectively, the system operates as a form of {}``heat engine''. The familiar textbook heat engine has the engine consume energy at rate $a$ to do {}``work" at rate $w$ to contribute to the potential of some outside agency while releasing waste {}``heat" at rate $a-w$  \cite{Zemanksy1997}. While the definition of work is clear for an industrial steam engine raising the gravitational potential of a steel beam, for example, the choice of what qualifies as work is quite subjective. In fact, all energy transfers $a$ act to increase the potential of something. Work is simply the raised potential of interest. Heat is the remainder.

With reference to Fig. \ref{fig:thermodynamic}, work is subjectively defined with respect to the internal energy potential $\Delta G$ of the interface separating the environment and the entity. Thus, through consumption of available energy at rate $a$, the value of $\Delta G$ evolves at rate $w=d\Delta G/dt$ with heat engine efficiency $\epsilon = w/a$. Meanwhile, heat is lost spontaneously to the colder surroundings at rate $a-w$. The Second Law of Thermodynamics requires that heat production $a-w>0$, in which case $\epsilon$ must be less than unity. Thus, the existence of a potential difference $\Delta G$ between the entity and its environment entails decay of the available potential of the universe as a whole (or equivalently, an increase in its entropy) \cite{Zemanksy1997}.  

The advantage of the above thermodynamic setup is that it allows for spontaneous evolution of the entity, and, as will be shown, it can be applied more specifically to the evolution of civilization. Because work is internal, a feedback loop causes the interface to exponentially grow or decay: the existence of $\Delta G$ requires energy consumption at rate $a = \alpha \Delta G$; in turn, this corresponds to work being done at rate $w=\epsilon a$, which then adds to the internal potential at rate $d\Delta G/dt = w$, closing the loop. If work increases the magnitude of $\Delta G$, then the interface separating the entity and its environment bootstraps itself to a higher level. Then the system as a whole evolves to higher levels of energy consumption $a$ through  \begin{equation}
\frac{da}{dt}=\alpha \frac{d(\Delta G)}{dt} = \alpha w = \alpha\epsilon a \equiv \eta a\label{eq:dadt}
\end{equation}
where $\eta$ is effectively a {}``rate of return'' representing the efficiency of the feedback on energy consumption $a$. 
Note that, perhaps counter-intuitively, higher energy efficiency $\epsilon$ corresponds to higher values of $\eta$,
and therefore more rapidly exponential evolution of energy consumption $a$ and heat production $a-w$.

In Appendix \ref{sec:thermo_details}, the nature of the feedback efficiency $\eta$ is defined more precisely. It is shown that the interface $\Delta G$ can be separated into $\breve{n}$ material units, each associated with the same potential energy at fixed temperature and pressure of $\Delta{\mu}$. $\Delta G$ results in a flux of material across the interface at rate $dn/dt = a/\Delta\mu$. If the net flow is from the environment to the entity, a portion of material that diffuses across the interface at rate $dn/dt$ then contributes to interface growth at rate $d\breve{n}/dt$. The feedback efficiency $\eta$ is the logarithmic form for this material rate of growth \begin{equation} 
\eta = \frac{d\ln{\breve{n}}}{dt}\label{eq:eta_and_n_1}\end{equation}

A concrete example that might be particularly easy to relate to is the growth of a young child. As an entity, the child 
consumes the accessible energy contained
in food from the environment in proportion to some measure of the child's size. This rate of consumption $a=\alpha\Delta G$ --
perhaps about 50 Watts -- enables the child to do {}``work'' $w=d\Delta G/d{t}$ with
energy efficiency $\epsilon=w/a$, incorporating the water and nutrients contained in food into its
structure in order to extend the material interface $\breve{n}$ separating it from its environment. The child maintains homeostasis because {}``heat'' can eventually radiate to space at rate $a-w$ at a relatively cold planetary blackbody temperature of about 255 K. Material waste is also produced once the useful chemical potential of the nutrition has been extracted, for example as carbohydrates are converted to exhaled CO$_{2}$.  Through a feedback loop, if $w>0$, the child and its energy consumption grow logarithmically at a rate $\eta=d\ln a/dt$. Of course, in an energy poor environment there might not be sufficient nutrition, in which case $w<0$ and the feedback efficiency $\eta$ is negative. But, assuming the child reaches adulthood, growth tends towards a balance between energy consumption and heat production, and $\eta$ tends to zero.

\section{Analog for the growth of civilization and its emissions}

The argument now is that the thermodynamic growth model described above, just as it can be applied to a child's growth, can also be extended to the human system in its entirety, as defined by civilization and its known environmental reservoirs. As with the child, an interface potential $\Delta G$ between civilization and its primary energy resources enables energy to be consumed at rate $a$. This allows work to be done with efficiency $\epsilon$ and at rate $w$ to grow the interface potential $\Delta G$ through incorporation of environmental matter (e.g., biomass and minerals). Simultaneously, through convection and radiation, heat is lost to space at rate $a-w$ at the planetary blackbody temperature. Also, unused material waste accumulates in the environment. 

Of course, for civilization, ``{}food'' includes the chemical and nuclear bonds in oil, coal and uranium, combined with mineral matter from the Earth's crust. These build structures that include not just human bones, flesh, blood and nerves, but cities, roads, shipping and telecommunications. However, insofar as the thermodynamics is concerned, the difference between the child and civilization is really only a matter of complexity and scale. In either case, as part of a single energy-consuming organism, all organism elements contribute to an interface with environmental reservoirs that  enables net available energy transfer to the organism at rate $a$. 

While a precise definition of civilization is arbitrary, civilization is most commonly quantified in purely fiscal terms. Thus, the goal here is to examine whether it is possible to link fiscal quantities to the more thermodynamic model defined above. To this end, an argument can be made that, if what physically distinguishes civilization from its environment is some thermodynamic potential $\Delta G$ at constant temperature and pressure, civilization implicitly assigns inflation-adjusted (or real) monetary value to what $\Delta G$ enables -- the total rate of energy consumption $a$.  To borrow a phrase, "money is power" because, if all current exothermic processes supporting civilization were to suddenly cease such that $a$ equalled zero, all civilization would become worthless; it would no longer be associated with
a non-equilibrium level of potential energy $\Delta G= a/\alpha$. Simply, there would be no definable material interface $\breve{n}$ between civilization and its environment.

As an example, the potential energy in oil combustion is valuable, but only to
the extent it that it can interact with the interface separating civilization from its environment. It has zero value if it burns
wastefully in the desert, and zero value in its unavailable chemical and nuclear bonds. From society's perspective, any societal element, whether living or synthetic, only has value to the extent it is able to operate in synergy with all other elements to define an interface with environmental available energy. An unavailable road from nowhere to nowhere is just pavement on the ground. But the same road between two cities is part of a larger organism that works collectively at net rate $w$ to grow access to the primary energy supplies that civilization requires.
The mathematical expression of the above argument is that global primary energy consumption $a$ is related to global value $C$ through a constant factor $\lambda$
\begin{equation}
a=\lambda C\label{eq:a}\end{equation}
Thus, the economic representation of the evolving heat engine given by Eq. \ref{eq:dadt} is \begin{equation}
\frac{dC}{dt}=\frac{1}{\lambda}\frac{da}{dt}=\frac{\alpha}{\lambda}\frac{d(\Delta G)}{dt} = \frac{\alpha}{\lambda} w = \frac{\eta}{\lambda}a\label{eq:P=dCdt}\end{equation}
or in purely economic terms \begin{equation}
 P \equiv \frac{dC}{dt}  = \eta C\label{eq:PetaC}\end{equation}
 where, $C$ (units value) grows through the real (inflation-adjusted) economic production rate $P$ (units currency per year). The thermodynamic feedback efficiency $\eta$ is an economic rate of return on $C$.

Eq. \ref{eq:P=dCdt} implies that real economic production $P$ is, perhaps rather intuitively, only a measure of thermodynamic work $w$ through coefficient ${\alpha/\lambda}$.  Expressed in integral form \begin{equation}
C\left(t\right)=\int_{0}^{t}P\left(t'\right)dt' = \frac{\alpha}{\lambda}\int_{0}^{t}w\left(t'\right)dt'\end{equation}
Thus, this growth model is a statement that the rate of return (or feedback efficiency) $\eta=\alpha\epsilon$ on economic value $C$ is a consequence of doing thermodynamic work $w$ with efficiency $\epsilon$ to grow the interface between civilization and environmental resources. By growing the interface, civilization is able to draw more energy $a$, and do more work $w$, thereby closing the loop. 

Now, returning to frameworks for CO${_2}$ emissions forecasts, the SRES definition for energy productivity $f=P/a$ can be understood in light of the above. If Eq. \ref{eq:dadt} is combined with  Eqs.  \ref{eq:P=dCdt}  and \ref{eq:PetaC}, this yields the basic relations  \begin{equation}
\eta = P/C = \alpha\epsilon = \lambda f \label {eq:eta}\end{equation}
Therefore, energy productivity $f=P/a$ is related to the heat engine thermodynamic efficiency $\epsilon=w/a$ and a fiscal expression for the feedback efficiency $\eta=P/C$ through the fixed, intrinsic quantities $\alpha$
and $\lambda$. 

\begin{figure}[htp]
\centerline{\includegraphics[width=0.8\textwidth]{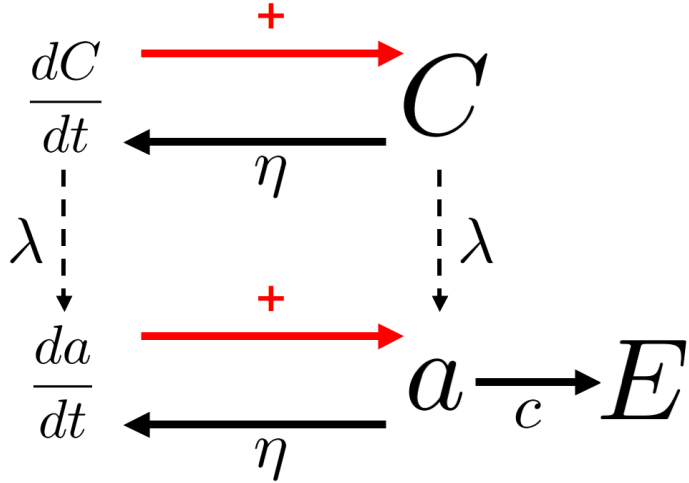}}
\caption{\label{fig:Schematic}Schematic extending Fig. \ref{fig:thermodynamic} to relate energy consumption $a$ by the human system to economic value $C$ and CO$_{2}$ emissions $E$. Black arrows point in the direction of the product, red arrows in the direction of the integral over time. Work is done at rate $w$ to enable energy consumption $a$ to grow at rate $da/dt = \eta a$, where $\eta$ is the feedback efficiency of a heat engine representing the system. The economy has a fixed relationship to energy consumption through $a = \lambda C$, where $C$ is civilization's historical accumulation of real (inflation-adjusted) economic production of economic value $P=dC/dt$ (units currency),  and $\lambda$ is an intrinsic constant of proportionality. Thus, CO$_2$ emissions are related to economic production through $E = \lambda c \int_{0}^{t}P\left(t'\right)dt'$, where $c$ is the carbon content of energy in the fuel supply.   }
\end{figure}

A schematic illustrating the economic growth model is shown in Fig. \ref{fig:Schematic}. A discussion of how it relates to more orthodox economic approaches is contained in Appendix \ref{sec:orthodoxy}, where it is shown how such traditional economic concepts as inflation, savings and capital depreciation can be interpreted within a thermodynamic context. A straightforward consequence of Eqs. \ref{eq:P=dCdt} and \ref{eq:PetaC} is that the rate of growth of the global economy obeys the simple relation \begin{equation}
\frac{d\ln P}{dt}=\eta+\frac{d\ln \eta}{dt} \label{eq:Pgrowth}\end{equation}

Interestingly, the approach is of identical mathematical form to one often used to successfully model growth of vegetation, where vegetative {} ``value" $C$ refers not to money but instead to biomass, and $P$ to the net primary productivity \cite{ThornleyJohnson1990,Montieth2000}. Presumably, biological organisms must also maintain a high potential interface with respect to their environment, enabling them to consume energy, produce heat and waste, and do work to incorporate the matter that enables them to grow \cite{Vermeij1995}. Thermodynamic laws are fully general. 

A difference between plants and civilization is that plant waste includes CO$_{2}$ that is recyclable, whereas the global economy creates most CO$_{2}$ from fossil-carbon, much of which accumulates in the atmosphere. From Fig. \ref{fig:Schematic},  CO$_{2}$ emissions can be represented simply through \begin{equation}
 E\left(t\right)=\lambda cC=\lambda c\int_{0}^{t}P\left(t'\right)dt'\label{eq:ElamcI} \end{equation} 
Present-day emissions are determined by past accumulation of real economic production and the current carbonization of the energy supply. Current emissions growth rates are given by 
\begin{equation}
\frac{d\ln E}{dt}=\eta+\frac{d\ln c}{dt} \label{eq:Egrowth}\end{equation}
Eq. \ref{eq:Egrowth} is more simple and physical than the expression for drivers in SRES forecasts given by Eq. \ref{eq:Kaya-prog}. 

\section{Evaluation}
The preceding discussion argues for a direct theoretical link between anthropogenic emissions and basic thermodynamics. But is the argument observationally supported? The expression for emissions growth, Eq. \ref{eq:Egrowth}, rests on the premise that there exists an intrinsic quantity $\lambda$ representing how the historical accumulation of economic production $C$ is supported by a rate of energy consumption $a$  (Eq. \ref{eq:a}). If $\lambda$ is not constant with time, then the thermodynamic framework is false.

\begin{figure}[htp]
\centerline{\includegraphics[width=0.8\textwidth]{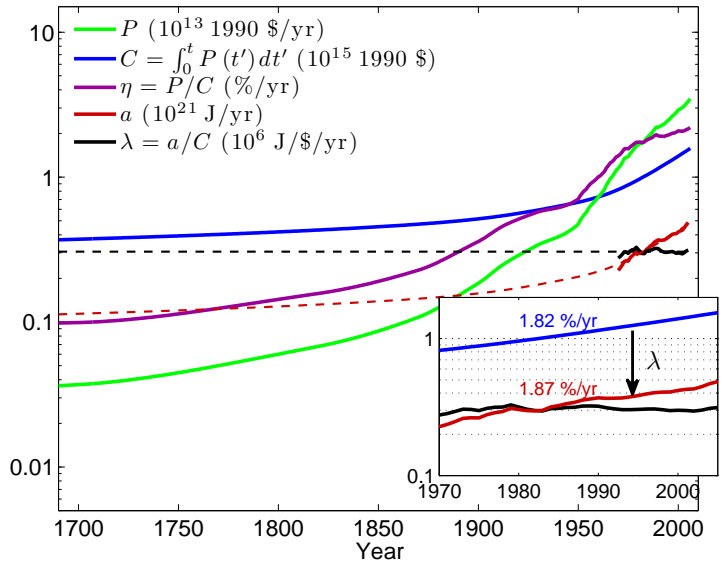}}
\caption{\label{fig:Ifigure}Estimates of gross world product $P$ in market
exchange rate, 1990 US dollars and economic value $C$, defined by $P=dC/dt$. Also shown are recent global primary energy consumption $a$, the ratio $\lambda = a/C$, and the feedback efficiency $\eta = P/C$. Dashed lines correspond to extrapolations based on assuming $\lambda = 9.7$ mW per 1990 US dollar.}
\end{figure}

I examine this proposition now using statistics for the combination of world energy production $a$ \cite{AER2006} and real global economic production $P$ \cite{UNstats} (expressed here in fixed 1990 US dollars) for the 36 year interval between 1970 to 2005 for which these statistics are currently available. The time series for accumulated global economic value $C=\int_{0}^{t}P\left(t'\right)dt'$ is estimated using sporadic calculations of $P$ that have been ascertained for select years over the past two millennia \cite{Maddison2003} in combination with more recent annual records \cite{UNstats} to
create a two-millennia yearly time-series in $P$ {[}See Appendix \ref{sec:time_series}]. Estimates of $P$ and $C$ and their ratio $\eta = P/C$ are shown in 
Fig. \ref{fig:Ifigure}. 

Figure \ref{fig:Ifigure} shows that, over a period between 1970 and 2005, the ratio $\lambda\left(t\right)=a/C$ maintained a nearly constant value of 0.306 exajoules per trillion 1990 US dollars per year, or alternatively $9.7$ milliwatts per 1990 dollar.  Corrected for autocorrelation in the time-series, the observational uncertainty at
the 95\% confidence level is just $\pm0.3$ milliwatts per 1990 dollar. The simplest interpretation is that this result supports the cornerstone hypothesis given by Eq. \ref{eq:a}: the historical accumulation of real economic value through real economic production is maintained by continuous primary energy consumption; the relationship between value and rates of energy consumption is a constant parameter. 

Of course it is possible that this observed result only holds over the 36-year period for which global energy consumption statistics are available, but it is expected theoretically; the period examined covers over half of total historical growth in $a$ and $C$, and two thirds of $P$; and, the observational uncertainty is small enough to plausibly reflect errors or noise in historical data. For example, new primary energy production (what has been measured) only reflects new primary energy consumption (what is theoretically relevant) in the average, not the instant.

\section{Drivers of emissions growth}

The existence of a fixed relationship between energy production and accumulated real economic production simplifies the number of drivers required for CO$_2$ emissions forecasts. To see how,  the SRES emissions growth equation Eq. \ref{eq:Kaya-prog} can be equated with the more thermodynamic expression given by Eq. \ref{eq:Egrowth}. Since both expressions rely on exogenous expressions for carbonization growth $d\ln c/dt$, this is effectively a comparison of expressions for growth in energy consumption $d\ln a/dt$ \begin{equation}
\frac{d\ln p}{dt}+\frac{d\ln g}{dt}-\frac{d\ln f}{dt}=\eta \equiv \lambda f\label{eq:Kaya-Garrett}\end{equation}
SRES models consider population $p$ and standard of living $g$ and energy productivity $f$ as the key {}``drivers'' of energy consumption growth, but the {}``driver'' concept can be misleading when, at a very basic level, feedback determines how $p$, $g$, and $f$ are inter-related. Eq. \ref{eq:Kaya-Garrett} demonstrates that growth in $p$ and $g$ is fundamentally constrained by the sum of the current state of the feedback efficiency $\eta\equiv\lambda f$ and its rate of change $d\ln{f}/dt$. Therefore, knowledge of the behavior of only one parameter, $f$, is required for forecasts of energy consumption growth, rather than each of $f$, $p$ and $g$.

So, perhaps surprisingly, changes in population and standard of living might best be considered as only a response to energy efficiency. As part of a heat engine, creating people and their lifestyles requires energy consumption. Doing so efficiently merely serves to bootstrap civilization into a more consumptive (and productive) state. Likely, society has traditionally praised energy efficiency gains for precisely this reason. As summarized by Eq. \ref{eq:dadt}, energy efficiency gains accelerate rather than slow energy consumption \cite{Jevons1865,Sorrell_UKERC2007}, contrary to what is commonly assumed \cite{PacalaSocolow2004} 
\begin{figure}[htp]
\begin{center}
\centerline{\includegraphics[width=0.8\textwidth]{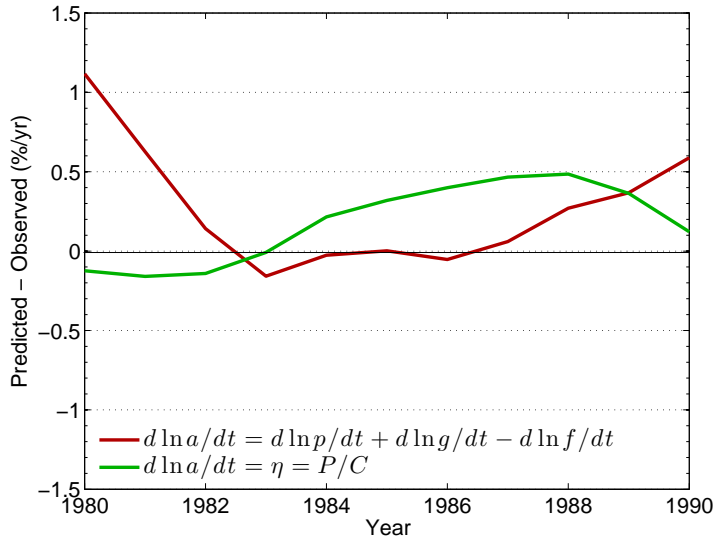}} \caption{\label{fig:comparison}Difference between predicted and actual growth
rates in global energy consumption $a$, based on predictions derived using persistence
in trends (red) and the zeroth-order model presented here (green).
Trend persistence is calculated using the prior ten
years as a basis; the zeroth-order model is based only on current year calculations
of the feedback efficiency $\eta$. Hind-casts are compared with actual observed rates for a period covering the following 15 years. }
\end{center} \end{figure}

This does not mean that consideration of population and standard of living in SRES models is invalid, of course, only that their evolution must be consistent with thermodynamic constraints (Eq. \ref{eq:Kaya-Garrett}). Fig. \ref{fig:comparison} shows a time-series comparing 15--year hind-casts with observations, evaluated for both sides of Eq. \ref{eq:Kaya-Garrett}, in each of the years 1980 to 1990. Hind-casts based on growth of $p$, $g$ and $f$ apply straight-forward persistence in trends from the prior 10 years, i.e., $d\ln a/dt = d\ln p/dt + d\ln g/dt- d\ln f/dt$. For comparison, hind-casts based on the zeroth-order thermodynamic expression for economic growth need only employ evaluations of the current--year state of $\eta = P/C$. 

Since both approaches reflect already realized thermodynamic constraints, both {}``persistence'' and a thermodynamic model give hind-casts that reproduce observed trends with comparable accuracy. Notably, the thermodynamic model provides a hind-cast for average growth between 1990 and 2005 that is within just 0.1 $\%$/yr of observed growth rates. A prior study found that this level of accuracy was only attained by a particular ``{}worst--case'' SRES model for this particular time period \cite{Raupach2007}. What is important here is that, for the purpose of future forecasts, the thermodynamic approach is accurate while being both simpler and more physical than using persistence or sophisticated SRES models. 

\section{Considerations for modeling future emissions scenarios}

An advantage of appealing to energy efficiency in forecasts of CO$_2$ emissions is that $\eta=\lambda f$ tends to vary rather slowly. Since 1970, growth of $\eta=P/C$ has climbed from 1.4 \% per year to 2.1 \% per year in 2005 (Fig. \ref{fig:Ifigure}), corresponding to an e-folding time-scale $\tau_\eta = 1/(d\ln\eta/dt)$ of approximately 100 years. Expressed in terms of time-series analysis, $\eta$ is highly {}``reddened'', because it is an integrator of $d\eta/dt$. Alternatively, and in more fiscal terms, global economic value $C$ (and hence energy consumption $a=\lambda C$) varies slowly because it is an integrator of economic production $P=dC/dt$ (Fig. \ref{fig:Ifigure}). The present and future are influenced by even the most distant past, and the past cannot be erased. 

The carbonization of the energy supply $c$, is changing even more slowly (see Appendix \ref{sec:rate_summary}) with a time scale of about 300 years. What this means is that future emissions rates for CO$_2$ are most strongly influenced by the current state of $\eta$. As a zeroth-order assumption, it is reasonable to assume persistence in  $\eta$, meaning that over time-scales much less than $\tau_\eta$, future emissions are unlikely to depart substantially from the recent growth rate of 2.1 \% per year. 

More accurate forecasts of energy consumption and CO$_2$ emissions rates will require an understanding of how $\eta$ itself evolves. Assuming $c$ is a constant, positive values of $d\ln\eta/dt$ imply super-exponential growth\footnote{Similar super-exponential growth behavior has been observed previously at a more local level, in the characteristics of cities \cite{Bettencourt2007}.} of CO$_{2}$ emissions $E$ \cite{Pielke2008}. The solution for Eq. \ref{eq:Egrowth}, starting at some time $t_i$, is \begin{equation}
{E}={E_i}\exp\left[\eta_{i}\tau_{\eta}\left(e^{\Delta{t}/\tau_{\eta}}-1\right)\right]\label{eq:asoln}
\end{equation}
Note, that growth condenses to the single exponential form in the limit of $\Delta{t} \ll \tau_{\eta}$. 

However, in the long-term, assuming persistence in even $d\ln\eta/dt$ is an over-simplification since $\tau_\eta$ itself evolves. History shows bursts in efficiency growth, notably around 1880 and 1950, perhaps when important new discoveries of energy reservoirs made the past less relevant (Fig. \ref{fig:Ifigure}). But, in both cases, the initial burst in $\eta$ eventually tapered. After 1950, the time-scale $\tau_\eta$, changed from just 30 years between 1950 and 1970, to 67 years between 1970 and 1990, and 120 years between 1990 and 2005. Plausibly, $1/\tau_\eta = d\ln\eta/dt$ will eventually cross zero and turn negative, implying sub-exponential growth in emissions $E$ (Eq. \ref{eq:asoln}).  

Unfortunately, if $-d\ln\eta/dt $ is ever greater than $\eta$ over the long term, while emissions growth may be significantly slowed, what is implied is a real global economy that is shrinking (Eq. \ref{eq:Pgrowth}). Robust multi-decadal forecasts of emissions $E$, and its relationship to economic production $P$, require a first principles thermodynamic model for how $\eta$ changes with time. Assuming 1880 and 1950 were indeed associated with discovery of new energy reservoirs, this would suggest the problem is fundamentally geological, and that higher-order moments of $\eta$ reflect rates of reservoir discovery and depletion. 

Understood thermodynamically, the transfer of energy at rate $a$ across the interface between energy reservoirs and civilization reflects a balance. On one hand, the transfer grows civilization, and increases the physical size of the interface $\Delta G$. At the same time, however, it depletes the reservoirs, and this decreases $\Delta G$. The sign and magnitude of the rate of work $w=d\Delta G/dt$, and therefore $\eta$, depends on the relative strengths of these two forces.

\section{Mitigation}

The premise behind mitigation is that there are ``{}drivers'' of emissions rates that can be meaningfully controlled through policy. As shown above, it is not clear that the driver concept is in fact meaningful. Rather, it appears that drivers in the Kaya Identity are merely a thermodynamic response to the current value of $\eta$. 

At least this appears to be true for population $p$ and standard of living $g$. It is not yet clear whether it applies to the current carbonization of the economy $c$. An interesting result that can be derived from Eq. \ref{eq:ElamcI}
using the values for $\lambda$ in Fig. \ref{fig:Ifigure} and $c$
in Fig. \ref{fig:trajectories}  (Appendix \ref{sec:rate_summary}) is that the {}``carbon footprint''
of civilization in recent decades reflects a simple relationship between
the rate of global carbon emissions $E$ and the accumulation over history of real global
value $C$. The coefficient is $\lambda c=$ 5.2$\pm$0.2 MtC per
year, per trillion 1990 US dollars of global economic value. 

To take the result further, Eq. \ref{eq:Egrowth} points towards a non-dimensional 
 number \begin{equation}
S=\frac{-d\ln c/dt}{\eta}\label{eq:decouple}\end{equation}
representing the relationship between the global economy's rate of de-carbonization,  $-d\ln{c}/dt$, and its rate of return, $\eta=\lambda f$. If $S\geq1$, $d\ln E/d{t}\leq0$, and emissions
are stabilized or declining. 

To reach stabilization, what is required is decarbonization that is at least as fast as the economy's rate of return. Taking the 2005 value for $\eta$
 of 2.1\% per year, stabilization of emissions would require an equivalent or greater
rate of decarbonization. 2.1\% of current annual energy
production corresponds to an annual addition of approximately
300 GW of new non-carbon emitting power capacity - approximately one
new nuclear power plant per day. 
  
 \section{Conclusions}
The physics incorporated into GCM representations of the land, oceans and atmosphere is required to adhere to universal thermodynamic laws. Ideally, the CO$_2$ emissions models meant for implementation in GCM projections of climate change should do so as well. Fortunately, it appears that appealing to thermodynamic principles may lead to a substantially constrained range of possible emissions scenarios. If civilization is considered at a global level, it turns out there is no explicit need to consider people or their lifestyles in order to forecast future energy consumption. At civilization's core there is a single constant factor, $\lambda = 9.7\pm0.3$ mW per inflation-adjusted 1990 dollar, that ties the global economy to simple physical principles. Viewed from this perspective, civilization evolves in a spontaneous feedback loop maintained only by energy consumption and incorporation of environmental matter. 

Because the current state of the system, by nature, is tied to its unchangeable past, it looks unlikely that there will be any substantial near-term departure from recently observed acceleration in CO$_2$ emission rates. For predictions over the longer term, however, what is required is thermodynamically based models for how rates of carbonization and energy efficiency evolve. To this end, these rates are almost certainly constrained by the size and availability of environmental resource reservoirs. Previously, such factors have been shown to be primary constraints in the evolution of species \cite{Vermeij1995,Vermeij2004}. Extending these principles to civilization, emissions models might be simplified further yet.

\begin{acknowledgements}
This work was enabled by financial support from the NOAA Office of Global Programs and a NASA
New Investigator Program award. The author is grateful for helpful feedback from Steven Sherwood, Geerat Vermeij, Claudio Holzner, Axel Kleidon, Clinton Schmidt, and Christopher Garrett. 
\end{acknowledgements}

\newpage

\appendix
\section*{Appendix}

\section{Material transfer\label{sec:thermo_details}}

To understand the details of the heat engine described in this article in a bit more detail, it is helpful to look more explicitly at what constitutes available energy and work for this case. The Gibbs energy potential of matter can be expressed as $\sum_{i}n_{i}\mu_{i}\left(T,p\right)$, where $n_{i}$ refers to the number of the species $i$ with specific chemical potential $\mu_{i}\left(T,p\right)$ \cite{Zemanksy1997,Job2006} (the {}``chemical'' potential, rather confusingly, can always be generalized where relevant to incorporate the potential exothermic energy in nuclear bonds). The interface separating the entity and its environment may be composed of matter in many forms. However, a simplifying argument can be made that the potential difference $\Delta G$ can be split up into $\breve{n}$ arbitrary units of matter, each unit carrying an identical available potential of $\Delta\mu\left(T_s,p\right)$  
 \begin{equation}
\Delta G = \breve{n}\Delta\mu\left(T_s, p\right) \label{eq:delta_G}
\end{equation} 
Likewise, the net flux of material between the environment and the entity at rate $d{n}/d{t}$ requires energy consumption by the system as a whole at rate \begin{equation}
a = \frac{dn}{dt}\Delta\mu\left(T_s,p\right) \label{eq:a_dndt}
\end{equation}
But, since $a=\alpha\Delta G$, it also holds that
 \begin{equation}
 a = \alpha\breve{n}\Delta\mu\left(T_s, p\right) \label{eq:a_nhat}
\end{equation}
Combined, Eqs. \ref{eq:a_dndt} and \ref{eq:a_nhat}, imply that the intensive quantity $\alpha = 1/\breve{n}\left(d{n}/d{t}\right)$ is determined by the particular physics relating the amount of high potential matter along the interface to the flux of matter across it.\footnote {It is straightforward to show that for the special case of Maxwellian diffusion along a concentration gradient to, for example, a cloud droplet or snow flake \cite{PruppacherKlett1997}, evolution of  $a$  and $\breve{n}$ is determined not by the surface area of the interface (as might initially seem more intuitive) but rather by a length dimension. In this case, $\alpha$ is determined by the product of the diffusivity of vapor in air and the area density of vapor at saturation.} .   

The evolution of energy consumption by the system,  $da/dt$, is related to its rate of doing work through $\alpha w$ (Eq. \ref{eq:dadt}), but more specifically to the interface's material growth. Since work is defined by $w = d\left(\Delta G\right)/dt$, and the interface temperature and pressure are fixed, it follows that the potential defining the interface between the entity and its environment evolves at rate \begin{equation} 
w = \frac{d\breve{n}}{dt}\Delta\mu\left(T_s,p\right)\label{eq:wequation}\end{equation}
Work is positive if the material interface grows. Expressed in terms of the rate of energy consumption, \begin{equation}
w = \frac{d\breve{n}}{d{n}}a  = \frac{1}{\alpha}\frac{ d\ln{\breve{n} } } {dt} {a} = \frac{1}{\alpha}\frac{da}{dt}\label{eq:wefficiency}\end{equation} 
Because the heat engine efficiency is given by $\epsilon = w/a$, this leads to the result that \begin{equation}
\epsilon = d\breve{n}/d{n} = \frac{1}{\alpha}\frac{d\ln{\breve{n}}}{dt} \label{eq:epsilon}
\end{equation} 
As a feedback loop, Eqs. \ref{eq:dadt} and \ref{eq:epsilon} can be combined to show that the feedback efficiency $\eta$ is \begin{equation} 
\eta = \frac{d\ln{\breve{n}}}{dt}\label{eq:eta_and_n}\end{equation}
Thus, $\eta$ expresses the logarithmic growth of the number of elements defining the interface between the entity and its environment. If the net flow is from the environment to the entity, the portion of material that diffuses across the interface and does not contribute to interface growth is returned to the environment as waste. 

\section{Comparison with traditional economic models\label{sec:orthodoxy}}

Economic studies normally separate production into two components: a fraction $s$ representing a savings, or investment;
and a fraction $\left(1-s\right)$ representing private and government
{}``consumption''. Models represent the nominal growth in {}``capital''
$K$ (units currency) as the difference between the portion $s$ of
production $P$ (units currency per time) that is a savings or investment,
and capital depreciation at rate $\gamma$\begin{equation}
\frac{dK}{dt}=\left(P-W\right)-\gamma K=sP-\gamma K\label{eq:dasKapital}\end{equation}
where individual and government consumption is represented by $W=\left(1-s\right)P$. 

In return, according to some functional form, labor $L$ (units worker
hours) employs capital $K$ (units currency) to generate further production
$P$ (units currency per time). For the sake of illustration, a commonly used representation
is the Cobb-Douglas production function \begin{equation}
P=AK^{\alpha}L^{1-\alpha}\label{eq:Cobb-Douglas}\end{equation}
where $A$, the {}``total factor productivity'', is a compensating
factor designed to account for any residual unaccounted for by $K$
and $L$. The exponent $\alpha$ is empirically determined. The Solow
Growth Model \cite{Solow1957} expresses the prognostic form for Eq.
\ref{eq:Cobb-Douglas} as \begin{equation}
\frac{d\ln P}{dt}=\frac{d\ln A}{dt}+\alpha\frac{d\ln K}{dt}+\left(1-\alpha\right)\frac{d\ln L}{dt}\label{eq:Solow}\end{equation}
 Commonly, the term $d\ln A/dt$ is interpreted to represent technological
progress.

There have been criticisms raised of the Solow Model because it makes
no explicit reference to natural resources \cite{Georgescu-Roegen,Ayres2003}. 
One suggested remedy has been to incorporate primary energy consumption
into Eq. \ref{eq:Cobb-Douglas} as a complement to labor or capital
\cite{Saunders1992,Saunders2000}, in which case \begin{equation}
P=\left(A_{K}K\right)^{\alpha}\left(A_{L}L\right)^{\beta}\left(A_{a}a\right)^{1-\alpha-\beta}\label{eq:Saunders_function}\end{equation}
where, again, $a$ is energy consumption, $\alpha$ and $\beta$ are
empirically determined, and the subscripts for $A$ refer to respective
technological progress. 

Now, by comparison, in the thermodynamic economic growth model introduced here, real (inflation-adjusted) economic production $P$ (units currency per time) and global value $C$ (units currency) are, respectively, fiscal representations of net thermodynamic work and the rate of consumption of available primary energy resources. The economic growth model described by Eqs \ref{eq:P=dCdt} and \ref{eq:PetaC} is given by the value production function 
\begin{equation}
P=\eta C\label{eq:PnuC}\end{equation}
where $\eta$ is the feedback efficiency representing a rate
of return due to thermodynamic work by the system on the system. The equation for growth of value is  \begin{equation}
dC/dt = P\label{eq:dCdt}\end{equation}
Note that while $C$ is analogous to capital $K$ in Eq. \ref{eq:dasKapital}, since $C= a/\lambda$, it is a more explicitly thermodynamic expression of value. 

So, to put the above in context of standard economic production functions, Eq. \ref{eq:PnuC}  can be considered
to be a simplification of Eq. \ref{eq:Saunders_function}.
The representation of economic value $C$ employed here is a substitution
of the combination of traditionally defined capital $K$ and labor
$L$ in Eq. \ref{eq:Saunders_function}, such that $\alpha=1$ and
$\beta=0$, and $A_{K}=\eta$. Alternatively, since $C$ is itself
only a monetary representation of the rate of primary energy consumption
$a$ through $a=\lambda C$, it could equally be stated that $\alpha=\beta=0$, and $A_{a}=\eta/\lambda$. 

Note that the thermodynamic production function (Eq. \ref{eq:PnuC}), unlike more standard
formulations (Eq. \ref{eq:Cobb-Douglas}), has the mathematical advantage of being dimensionally self-consistent,
as it does not need to appeal to non-integer exponents $\alpha$ and $\beta$
of dimensional terms (such as $L$ and $K$), as fitted to a specific
set of circumstances, and with no certain application to different
economic regimes. 

It might be argued, however, that the model introduced here fails by leaving no room for either consumption $W$ or depreciation $\gamma K$, two central components of the standard economic growth equation (Eq. \ref{eq:dasKapital}). Offhand, this seems reasonable because, certainly, some portion of economic production must be consumed, at least in order to maintain economic capital against depreciation or decay: buildings crumble; bodies must
be maintained; old technology becomes obsolete; as does past acquisition
of human skills and knowledge.

But these concerns can be resolved once it is recognized that the equations derived for this study are
intended to apply only to real, inflation-adjusted production $P$,
and not nominal production $\hat{P}$. To demonstrate, assume inflation
is positive, in which case nominal value $\hat{C}$ grows faster
than real value by some fractional rate of real value, $\gamma$
\begin{equation}
\frac{d\hat{C}}{dt}=\frac{dC}{dt}+\gamma C\label{eq:nominal-C-growth}\end{equation}
Since it has been argued here that $dC/dt=P$, this leads to \begin{equation}
\frac{dC}{dt}=\hat{P}-\gamma C\label{eq:real-C-growth}\end{equation}
in which case, the source of real value is nominal production, and
the corresponding sink for real value occurs at rate $\gamma$.
So, in fact, Eq. \ref{eq:real-C-growth} illustrates that Eq. \ref{eq:dCdt} does account for depreciation
through the term $\gamma C$, and is thus similar to the depreciation
term $\gamma K$ in the standard growth equation for capital (Eq.
\ref{eq:dasKapital}). While depreciation is implicit when the growth
equations are expressed in real, inflation-adjusted terms, depreciation
is explicit when they are expressed in nominal terms. 

In fact, it is interesting to see what the value decay rate $\gamma$ represents.
Again, because $dC/dt=P$, this means Eq. \ref{eq:dCdt}
leads to the statement $\hat{P}-P=\gamma C$. Alternatively, when
nominal production is expressed in energy consumption co-ordinates
through substitution of the expression $a=\lambda C$\begin{equation}
\hat{P}-P=\gamma C=\frac{\gamma}{\lambda}a\label{eq:nominal-P}\end{equation}
Compare this to an equivalent expression derived for real production
$P=({\eta}/{\lambda})a$ (Eq. \ref{eq:P=dCdt}). The implication here is that economic inflation, the difference between nominal and real production, is a consequence of the spontaneous decay or depreciation of total economic value $C$ at rate $\gamma C$. Put another way, since $P=\eta C$, the ratio $\gamma/(\gamma+\eta)$
is the fraction of nominal production $\hat{P}$ that, unlike $P$, does not return
itself as a real addition to total value $C$. In thermodynamic terms, value depreciation
 $\gamma C$ is an energy barrier that must first be crossed for real production to occur. If it is not, the perspective of civilization is that nominal production may be positive, but real production is negative. Net work $w$ is done on civilization by the environment, rather than the reverse. Whenever this occurs, the interface between civilization and its environment $\Delta G$ decays.

There is also a consumption term in the traditional expression for capital growth Eq. \ref{eq:dasKapital} that is not present in the thermodynamic expression for total value growth  Eq. \ref{eq:dCdt}.  As it is normally defined, consumption is the portion of economic production that does not represent an investment or savings in traditional representations of capital $K$. By contrast, in the thermodynamic model, effectively all real production is an ``{}investment'' in total economic value $C$. While a portion of nominal production or nominal work may merely serve to offset depreciation of $C$ as described above, all of the remainder adds to the total. Real production is net production.

To illustrate, the construction of coal mines and power plants clearly represents
an investment in economic value in either framework. A less obvious, although functionally
equivalent example, is food consumption. In standard representations, food would
be {}``consumed'' by households and not contribute to their value. However, the available chemical
potential in food consumption $dn/dt\Delta\mu$ (Eq. \ref{eq:a_dndt}) also maintains and improves
that household's capacity to further consume energy and do work by
supporting its internal potential energy $\Delta G$ (Eq. \ref{fig:thermodynamic}). Of course, the consumption
of an ordinary sandwich may only offset a body and mind against
decay from {}``heat'' loss, maintaining its internal potential such that it can continue to consume
energy at the same rate it has in the past (in which case the real
production rate $P$ and net work rate $w$ is zero since $\hat{P}=\gamma C$). The added value
of a really good, if more expensive, sandwich is its capacity to facilitate
real production and new energy consumption above and beyond decay
(in which case real production is greater than zero and $\hat{P}>\gamma C$).
The addition to total global value $C$ (and internal potential $\Delta G$) may derive from a heightened sense
of personal well-being and an increased desire to productively interact with the rest of civilization in order to
afford such sandwiches. 

It is worth noting that a primary conclusion of this paper, that feedback loops in the economic system mean that  increases in energy efficiency correspond to greater energy consumption, has been reached previously by some economists, albeit in a less explicitly physical form than presented here \cite{Ayres2003,Saunders1992,Saunders2000,Khazzoom1980,Brookes1990,Alcott2005,Polimeni2006,Dimitropoulos2007,Herring2007}. Although the concept was first introduced by W. Stanley Jevons over a century ago \cite{Jevons1865}, the extent of energy efficiency {}``rebound'' or {}``backfire'' remains disputed \cite{Sorrell_UKERC2007}, with no consensus on how it should be quantified on the global scales relevant to forecasts of climate change from anthropogenic CO$_2$ .
\section{Materials and methods for time series estimates \label{sec:time_series}}

US Department of Energy statistics for global primary energy production
\cite{AER2006} include fossil fuel, hydroelectric, nuclear, geothermal,
wind, solar, and biomass sources. It is assumed here that production
and consumption rates are, at least on average, equivalent. United
Nations time series for world economic production \cite{UNstats}
represent the total gross domestic product of all countries, adjusted
for inflation and market exchange rates to fixed 1990 US dollars.
Statistics for CO$_{2}$ emissions are obtained from the Carbon Dioxide
Information Analysis Center \cite{Marland2007}. Rather than looking at nations or sectors, only global quantities are considered here because, at this level, atmospheric CO$_2$ is well-mixed, and international markets make details in economic trade unimportant.

Gross World Product estimates in 1990 market
exchange rate dollars are available for each year since 1970 \cite{UNstats}.
Long-term but intermittent historical estimates are available for
the years 1 to 1992 CE \cite{Maddison2003}. The latter data set
is expressed in Geary Khamis purchasing power parity (PPP) 1990 US
dollars. In general, the motivation for expressing valuation in PPP instead
of exchange rate dollars is to account for disparities in product
valuation that exist between countries. In PPP dollars, product valuation
is equalized according to its apparent contribution to standard of
living. Countries with a low standard of living tend to have a relatively
high gross domestic product when expressed in PPP rather than market
exchange rate dollars because equivalent products and services tend
to be less expensive. 

However, because the focus of this study is
energy production and associated CO$_{2}$ emissions, rather than
national standard of living, it is historical records of market exchange
rate valuations that are used. Exchange rate measures of production
$P$ are assumed to most accurately reflect the total energy costs
associated with manifesting products and services in the respective
nations where they are consumed. 

To account for any discrepancy between PPP and exchange rate estimates
in historical records for economic production $P$, market exchange
rate data from 1970 onwards is used to devise a time-dependent correction
factor $\pi$ to be applied to PPP records such that $\pi=\mathrm{PPP}/\mathrm{exchange}\,\mathrm{rate}$. For the period 1970 to 1992, during which both PPP and market
exchange rate estimates of $P$ are available, the fitted value for
$\pi$ is $\pi=1+0.258\exp\left[\left(t-1998\right)/73\right]$. This
correction factor can be extrapolated and applied to all PPP data
between the years 0 and 1969. For the period from 1970 onwards, measured
exchange rate values are used. Because the historical estimates of
$P$ in PPP dollars are increasingly sparse with distance back in
time (e.g. there are only three data points for the period 1 to 1500
CE), the corrected dataset for $P$ is mapped to a yearly distribution
using a cubic spline fit. 

Estimates of economic value $C$ represent an accumulation of economic production $P$ over time since
1 CE, i.e., $C\left(t\right)=C\left(1\right)+\int_{1}^{t}P\left(t'\right)dt'$.
To estimate a value for $C\left(1\right)$, it is assumed that the
ratio of population to economic value in 1 CE. is equivalent to
the average value between 1 CE and the threshold of the industrial
revolution circa 1700 CE. From historical population statistics \cite{Maddison2003},
the associated iterative solution for $C\left(1\right)$ is 120 trillion
1990 U.S. dollars. For comparison, the estimated value of $C$ in
2005 CE is 1580 trillion 1990 US dollars (Fig. 3). Although, off-hand,
this value for $C\left(1\right)$ seems surprisingly high, it is still
very small compared to current day values, so the derived value of
$\lambda$ presented in this paper is relatively insensitive to errors
in its estimate.

\section{Summary of observed growth rates between 1970 and 2005\label{sec:rate_summary}}

\begin{figure*}[htp]
\includegraphics[width=0.5\textwidth]{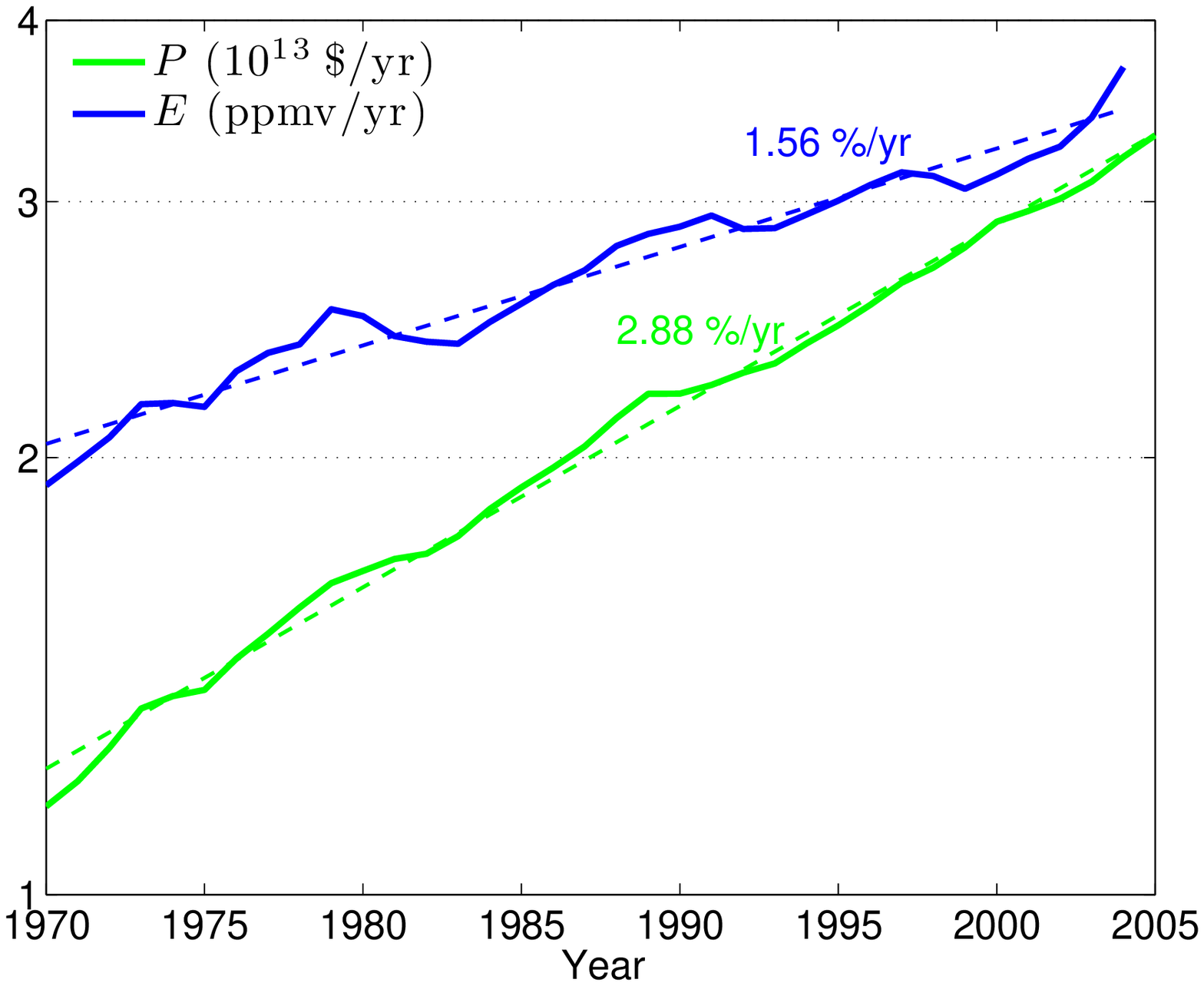}
\includegraphics[width=0.5\textwidth]{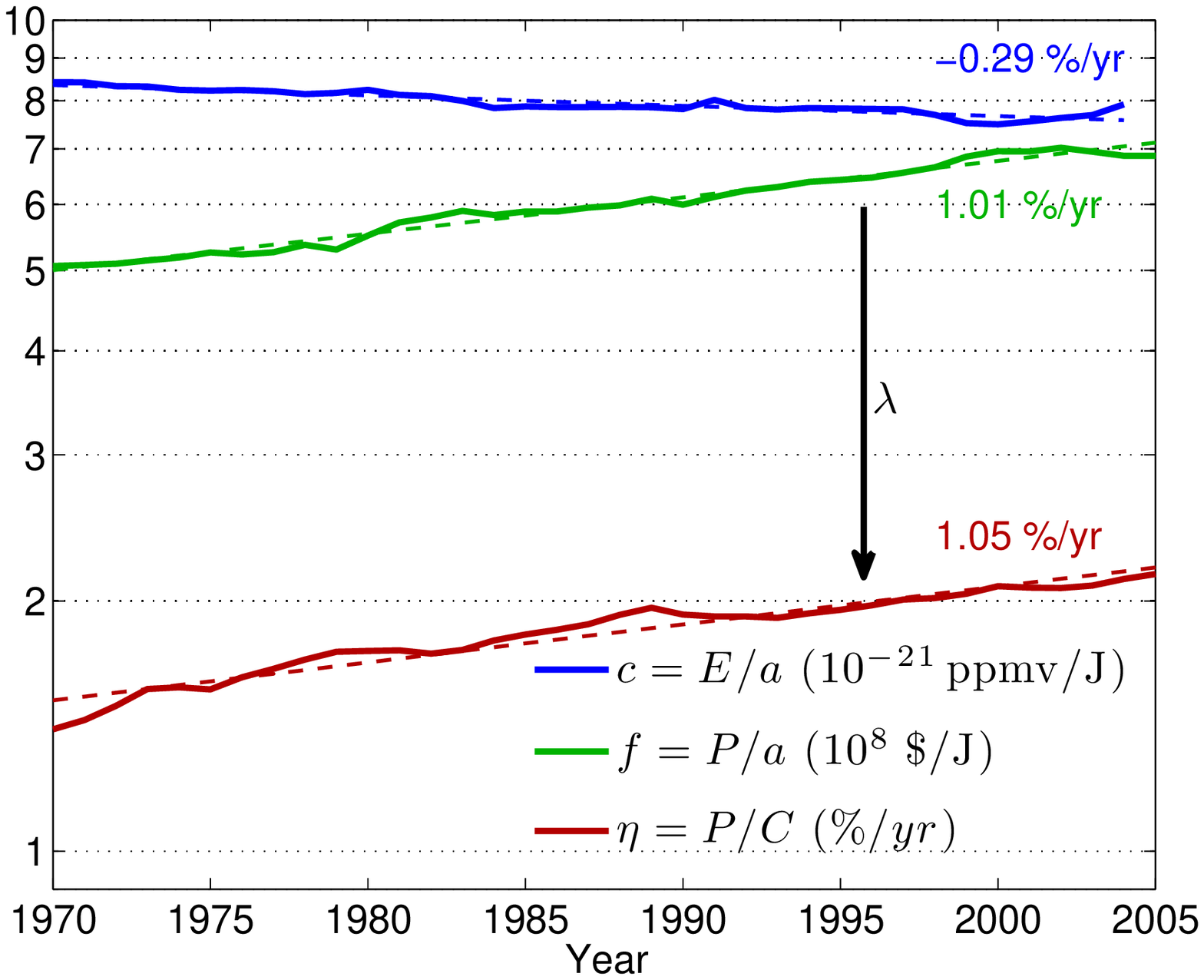}

\caption{\label{fig:trajectories}For the period 1970 to 2005, trajectories
in real global world production $P$ and carbon dioxide emissions
$E$ (left), and feedback efficiency $\eta = P/C$
and the carbon dioxide emission intensity of energy $c=E/a$ (right).
Here, $c$ represents the increase in atmospheric concentrations of
CO$_{2}$ $E$, per unit primary energy consumption $a$, that would
be expected in a well-mixed atmosphere in the absence of terrestrial
sink and source terms (1 ppmv CO$_{2}$ = 2.13 Gt emitted carbon \cite{Trenberth1981}).
Dashed lines represent a least-squares first-order fit. Theoretical relationships between parameters are summarized in Table \ref{tab:Summary-of-average}}

\end{figure*}

A summary of observed growth rates in global world real production $P$, carbon dioxide emission rates $E$, feedback efficiency $\eta$ and carbon dioxide emission intensity $c$ between 1970 and 2005 is provided in Fig. \ref{fig:trajectories} and in Table \ref{tab:Summary-of-average}, along with relevant equations based on the thermodynamic model described here.  

\begin{table}[htp]
\caption{\label{tab:Summary-of-average}Mean observed and modeled
 quantities for the period 1970 to 2005. }
\begin{tabular}{cccc}
\hline 
Parameter  & Functional dependence & Observed mean & Model mean\tabularnewline
\hline
energy efficiency growth & $d\ln\eta/dt$ & 1.05 \%/yr & -\tabularnewline
carbonization growth  & $d\ln c/dt$ & -0.29 \%/yr & -\tabularnewline
feedback efficiency & $\eta=\eta_{i}\exp\left(\frac{d\ln\eta}{dt}t\right)$  & 1.84 \%/yr & -\tabularnewline
energy productivity growth & $d\ln{f}/dt$ & 1.01 \%/yr & 1.05 \%/yr \tabularnewline
energy consumption growth & $d\ln a/dt=\eta$  & 1.87 \%/yr & 1.84 \%/yr\tabularnewline
economic value growth & $d\ln C/dt=\eta$  & 1.82 \%/yr & 1.84 \%/yr\tabularnewline
economic production growth & $d\ln P/dt=\eta+d\ln\eta/dt$  & 2.88 \%/yr & 2.89 \%/yr\tabularnewline
CO$_{2}$ emissions growth & $d\ln E/dt=\eta+d\ln c/dt$  & 1.56 \%/yr & 1.55\%/yr\tabularnewline
\hline
\end{tabular}
\end{table}

\newpage

\bibliographystyle{spmpsci}

\end{document}